\documentclass[12pt,a4paper,dvips]{article}
\usepackage{epsfig,wrapfig} 
\renewcommand{\section}[1]{\vspace{6pt} \noindent\mbox{#1} \newline \noindent}
\renewcommand{\subsection}[1]{\vspace{6pt} \noindent\mbox{\underline{#1}} 
\newline \noindent}
\renewcommand{\subsubsection}[1]{\vspace{6pt} \noindent\mbox{\underline{#1}}
\noindent}

\newfont{\sansb}{cmssbx10}
\newfont{\sans}{cmss10}

\def\Journal#1#2#3#4{{#1} {\bf #2}, #3 (#4)}

\def\AJL{{\em Ap.J. Lett.}}

\begin{document}

{\center \ {\large DETECTION OF VHE GAMMA-RAYS FROM MRK 501 
WITH THE CAT IMAGING TELESCOPE}
\vspace{30pt}\\

A.~Barrau$^5$, R.~Bazer-Bachi$^2$, H.~Cabot$^7$, L.M.~Chounet$^4$, G.~Debiais$^7$,
B.~Degrange$^4$, J.P.~Dezalay$^2$, A.~Djannati-Ata\"{\i}$^5$, 
D.~Dumora$^1$, P.~Espigat$^3$, B.~Fabre$^7$, P.~Fleury$^4$, G.~Fontaine$^4$, R.~George$^5$,
C.~Ghesqui\`ere$^3$, P.~Goret$^6$, C.~Gouiffes$^6$, I.A.~Grenier$^{6,9}$, L.~Iacoucci$^4$,
S.~Le Bohec$^3$, I.~Malet$^2$, C.~Meynadier$^7$, F.~Munz$^8$, T.A.~Palfrey$^{10}$, E.~Par\'e$^4$, 
Y.~Pons$^5$, M.~Punch$^3$, J.~Qu\'ebert$^1$, K.~Ragan$^1$, C.~Renault$^{6,9}$, 
M.~Rivoal$^5$, L.~Rob$^8$, P.~Schovanek$^{11}$, D.~Smith$^1$,  J.P.~Tavernet$^5$
and J.~Vrana$^{4 \dag}$ \vspace{6pt}\\
{\it $^1$Centre d'Etudes Nucl\'eaire de Bordeaux-Gradignan$^{\ast}$, France \\
$^2$Centre d'Etudes Spatiales des Rayonnements$^{\ddagger}$, Toulouse, France \\ 
$^3$Laboratoire de Physique Corpusculaire$^{\ast}$, Coll\`ege de France, Paris, France \\
$^4$Laboratoire de Physique Nucl\'eaire de Haute Energie$^{\ast}$, Ecole Polytechnique, Palaiseau, France \\
$^5$Laboratoire de Physique Nucl\'eaire de Haute Energie$^{\ast}$, Universit\'es de Paris VI et VII, France \\
$^6$Service d'Astrophysique$^{\#}$, Centre d'Etudes de Saclay, France \\
$^7$Groupe de Physique Fondamentale$^{\ast}$, Universit\'e de Perpignan, France \\
$^8$Nuclear Center, Charles University, Prague, Czech Republic \\
$^9$Universit\'e Paris VII \\
$^{10}$Department of Physics, Purdue University, Lafayette, IN 47907,
U.S.A \\
$^{11}$JLO Ac. Sci. \& Palacky University, Olomouc, Czech Republic \\
$^{\ast}$IN2P3/CNRS \\
$^{\ddagger}$INSU/CNRS \\
$^{\#}$DAPNIA/CEA \\
$^{\dag}$Deceased \\
}}

 {\vspace{12pt}}
{\center ABSTRACT\\}
The CAT imaging telescope on the site on the former solar plant Th\'emis has been
observing $\gamma$-rays from Mrk501 above 220 GeV in March and April 1997. This
source is shown to be highly variable and the light curve is presented. The detected
$\gamma$-ray rate for the most intense flare is in excess of 10 per minute.

\setlength{\parindent}{1cm}
\section{INTRODUCTION}
The CAT imaging telescope was recently built on the site of the former solar plant "Th\'emis"
(French Pyr\'en\'ees).  
The detector has a trigger threshold for $\gamma$-rays
of  $\sim$220~GeV near the zenith, with excellent source-location and
energy-resolution capabilities. The detector has been operational since
October, 1996.
\noindent
The active galactic nucleus (AGN) Markarian 501 was the second extragalactic
source detected in the VHE domain , by the Whipple group (Quinn et
al., 1996) with a flux level of a tenth that of the Crab.  In 1996 observations,
some flaring activity was noted (Weekes, 1997).

\section{OBSERVATIONS}
The CAT imaging Cherenkov telescope has observed Mrk 501 in March and 
April of 1997. The total observing time on-source is 50 h,
with 18 h of data off-source, i.e. on a control region of the 
sky following the same trajectory as the source.  Data were taken under
clear-sky, moonless conditions.  At the latitude of the Th\'emis site,
Mrk501 passes close to the zenith, thus providing a low threshold for
atmospheric Cherenkov detection.

\section{DATA-ANALYSIS}
The data were analysed using both the standard moment-based analysis and the
model-based maximum likelihood method developed by the CAT group (Le Bohec,
1996) which takes advantage  of the  fine resolution of the CAT imaging
camera.  The latter method has the advantage of providing the point of origin
on the sky (in two  dimensions) of each individual $\gamma$-ray, and gives an
estimate of the $\gamma$ energy. 

\section{RESULTS}
For the two months of observation, a strong signal was seen from  Markarian
501.  For the moment-based analysis, Figure 1 shows the  distribution of the
orientation angle ($\alpha$) for those events which pass the shape cuts, for all
of the on-source observations at a zenith angle less than 30$^{\circ}$.
 It can be seen that there is a strong signal
at small $\alpha$ values, as expected for a point source, and that the control
region shows no excess. 
The model-based maximum likelihood method also shows a strong signal, as shown
in Figure 2; in this case $\alpha$ is defined as the angle at the image
centre between the source position and the estimated point of origin of the
$\gamma$-ray.  In Figure 3, the 2-dimensional binned distribution of the 
estimated $\gamma$-ray origins for the on-source data minus the off-source 
data is plotted.  The position of maximum emission is clearly seen to be at
the position of Markarian 501.
\noindent
Given that this source is known to have flaring episodes, the night-to-night
variation in the event rate was investigated, using a moment-based analysis.
The cuts efficiency for gamma
rays is of the order of 40\%. Figure 4 shows the resulting
nightly $\gamma$-ray rate, for events passing all cuts. 
  The source is seen to exhibit rapid variations of an order of magnitude
, in contrast to the Crab Nebula which was seen to 
be steady from October 1996 to January 1997.  
At the same site in April 1997, the Themistocle array (Djannati-Atai et al., 1995) has simultaneously observed
flaring activity from Mrk 501 above about 1.5 TeV; we are currently analysing a sample
of common events.

\section{CONCLUSIONS}
In spring of this year, Markarian 501 has exhibited intense and highly-variable
flaring activity, becoming at times the brightest known source in the
VHE $\gamma$-ray sky. We are now investigating the spectral shape of Markarian 501.

\section{REFERENCES}
\setlength{\parindent}{-5mm}
\begin{list}{}{\topsep 0pt \partopsep 0pt \itemsep 0pt \leftmargin 5mm
\parsep 0pt \itemindent -5mm}
\vspace{-15pt}
\item Djannati-Atai, A. Proc. 24th ICRC, 2, 315 (1995).
\item Le Bohec, S., ``Conception et R\'ealisation d'un T\'elescope \`a Effet
Tcherenkov Atmosph\'erique pour l'Astronomie Gamma de 100 GeV \`a 10 TeV'',
Ph.D. thesis (1996)
\item Quinn, J., Akerlof, C.W., Biller, S., {\it et al.}, ``Detection of Gamma
Rays with E $>$ 300~GeV from Markarian 501'', \Journal{\AJL}{456}{L83}{1996}.
\item Weekes, T.C., ``Recent results from Whipple'', Moriond (1997)

\end{list}

\begin{figure}[hb]
 \begin{center}
   \mbox{\epsfig{file=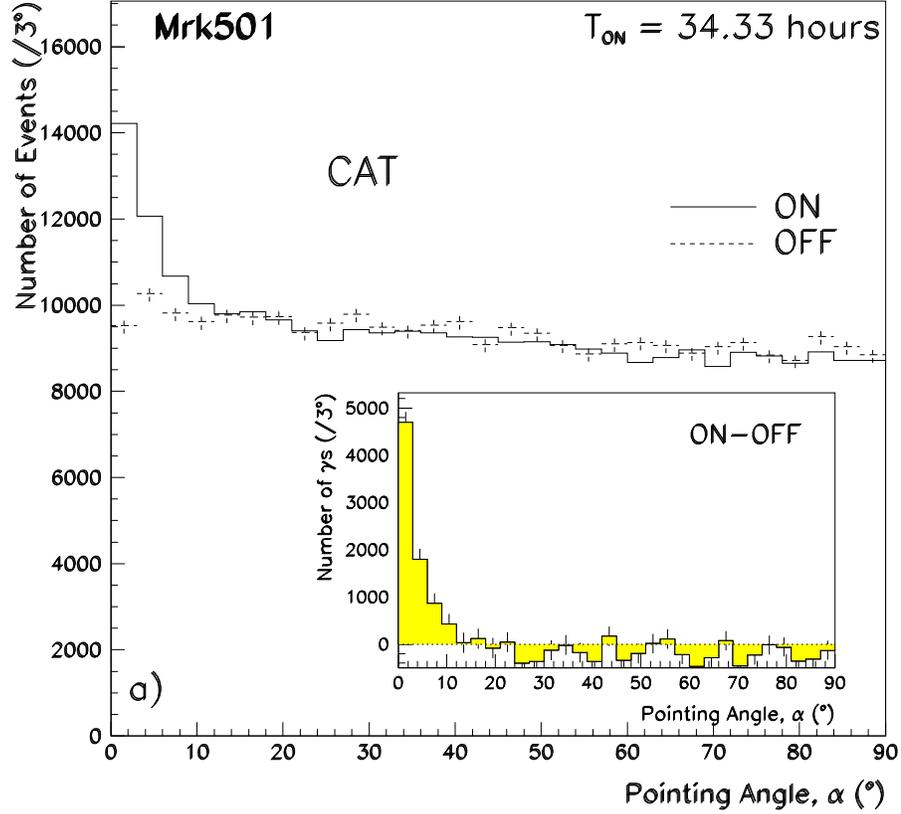,height=10.5cm}}
 \end{center}
 \caption{The $\alpha$-plot of Mrk501 using the moment-based analysis. The bin width is
 3$^{\circ}$ and the zenith angle for this data set is less than
 30$^{\circ}$.}
\end{figure}

\begin{figure}[ht]
 \begin{center}
   \mbox{\epsfig{file=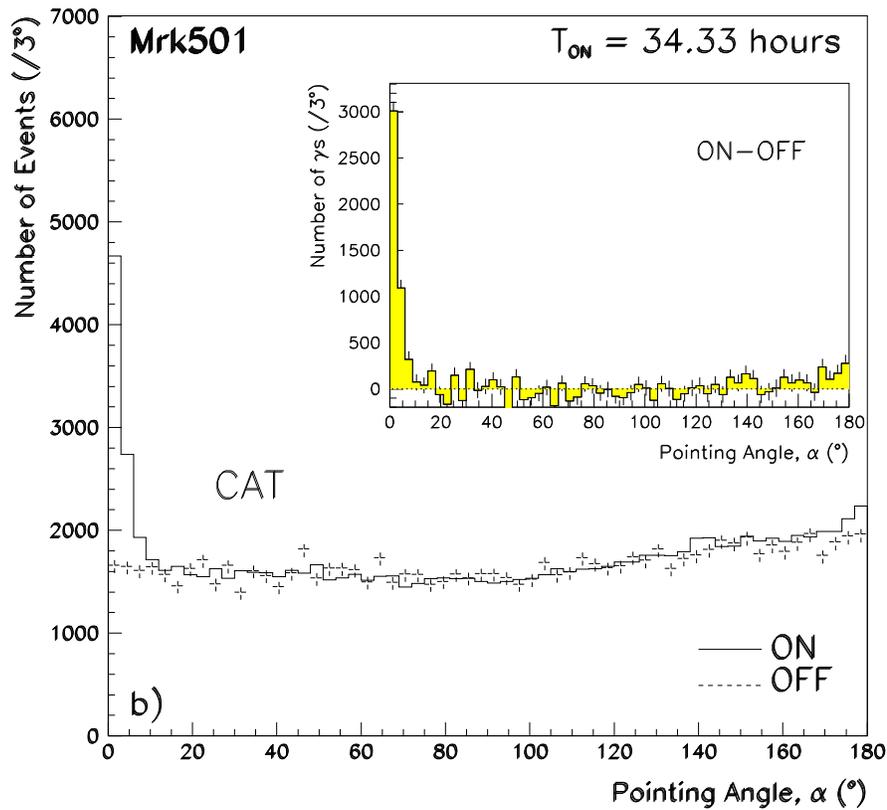,height=10.5cm}}
 \end{center}
 \caption{The $\alpha$-plot of Mrk501 using the model-based analysis of 
 LeBohec (1996). The bin width is
 3$^{\circ}$ and the zenith angle for this data set is less than
 30$^{\circ}$.}
\end{figure}

\begin{figure}[hb]
 \begin{center}
   \mbox{\epsfig{file=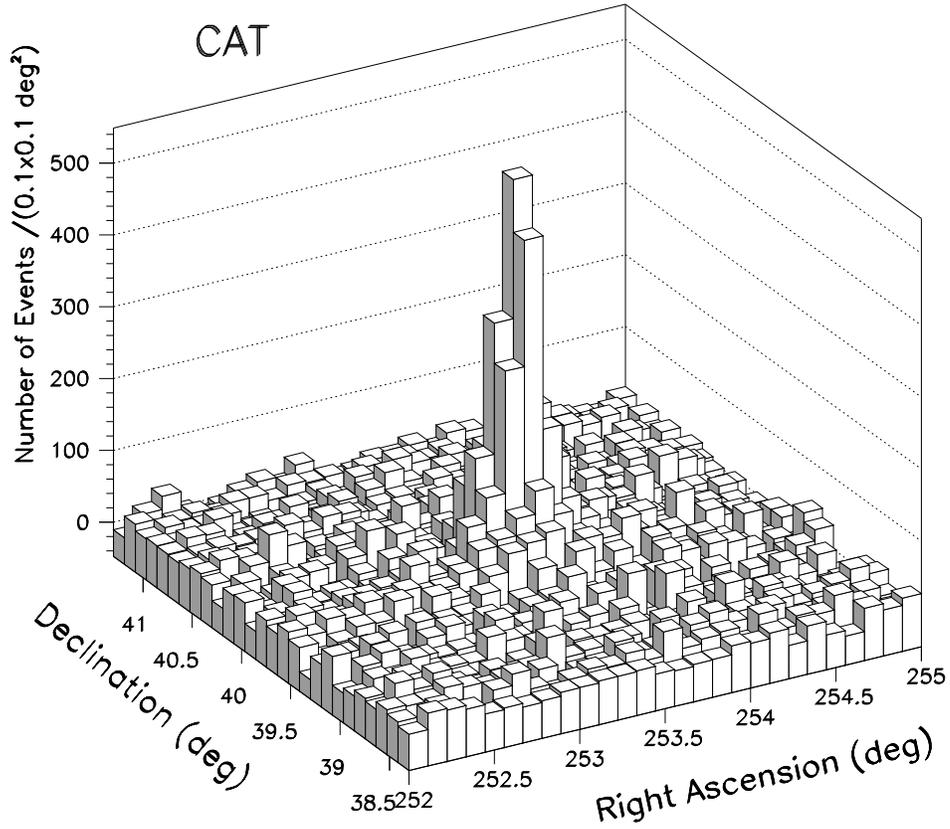,height=10.5cm}}
 \end{center}
 \caption{The ON-OFF plot of Mrk501 in RA,DEC coordinates as obtained with the method 
 of LeBohec (1996). The bin size is 0.1$\times$0.1 deg$^2$.}
\end{figure}

\begin{figure}[t]
 \begin{center}
   \mbox{\epsfig{file=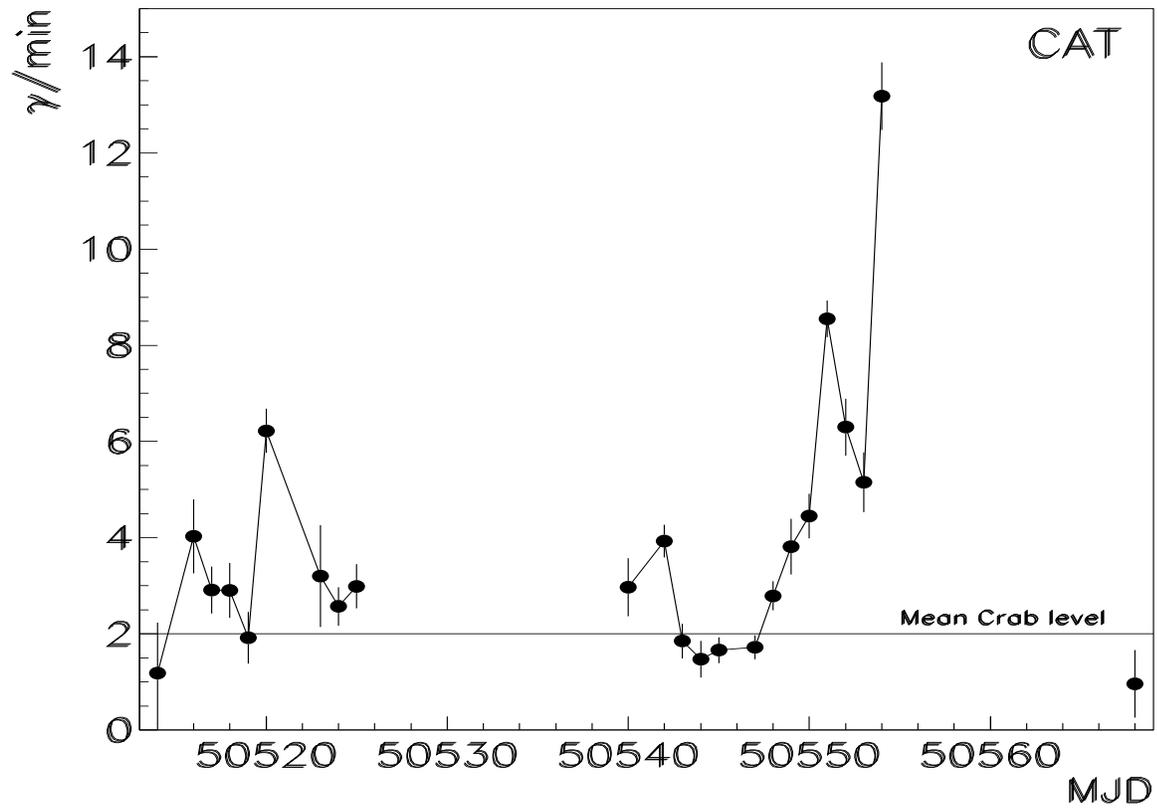,height=12cm,width=16cm}}
 \end{center}
 \caption{The $\gamma$-ray rate from Mrk501 on a night to night basis after shape and
 orientation cuts. The horizontal line shows the
 average Crab rate.}
\end{figure}

\end{document}